\pdfoutput=1
\documentclass[10pt]{article}

\usepackage[utf8]{inputenc}
\usepackage[T1]{fontenc}
\usepackage[eandd,preprint]{neurips_2026}
\makeatletter
\renewcommand{\@noticestring}{Preprint. Accepted at NeurIPS 2026 (Track on Evaluations and Datasets).}
\makeatother
\usepackage{amsmath,amssymb,amsthm}
\usepackage{booktabs}
\usepackage{graphicx}
\usepackage{url}
\usepackage{xcolor}
\usepackage{hyperref}
\usepackage{microtype}
\usepackage{enumitem}
\usepackage{caption}
\usepackage{subcaption}
\usepackage{multirow}
\usepackage{array}
\usepackage{tabularx}
\usepackage{float}
\usepackage{listings}

\hypersetup{colorlinks=true,linkcolor=blue!60!black,citecolor=blue!60!black,urlcolor=blue!60!black}

\newcommand{\bench}{CTE-Bench}

\title{\bench: Counterfactual Trace Evaluation for\\Stateful Software Simulators}

\author{%
  Xinran Zhang\\
  Arga Labs\\
  UC Berkeley\\
  \texttt{xinran@argalabs.com}
}
\date{}

\begin{document}
\maketitle

\begin{abstract}
Coding agents change running software: they patch a
service's code or overwrite its stored state, and then act on their own
expectation of how the service will respond afterwards. A wrong expectation
may surface only several calls later. Function-level code-execution
benchmarks omit persistent service state, and agent benchmarks score
the actions an agent takes or the final state it reaches. We
introduce \bench{}, which measures whether a model can predict
how an intervention changes a stateful service's future behavior, without
asking it to choose actions. Each scenario gives the model Python service
code, the calls and responses observed before the intervention, the
intervention itself (a source edit or a state overwrite), and 40 fixed
future calls; the model predicts every future response, and predictions
are checked by executing the service. Three memory protocols control
whether the model sees the correct earlier responses, none of them, or
its own earlier predictions. \bench{}-Core-v1 contains 255 scenarios over
six deterministic Python services, giving 10,200 predictions per model. The main score is effect-step value match (VM): exact
response equality on the 2,476 future calls whose response the
intervention changes. With correct earlier responses revealed, four API-hosted models
(DeepSeek~V4-Flash, Kimi~K2.5, Qwen3.6-35B-A3B, and Claude~Sonnet~4.6)
reach 54.3\%--61.7\% effect-step VM. Hiding those responses lowers effect-step VM
to 23.2\%--29.0\%; conditioning on self-generated predictions gives
24.8\%--33.2\%, and at most 1.2\% of scenarios are predicted exactly end
to end. Current models thus track intervention effects mainly when
correct feedback is supplied, and their errors compound over a rollout.
We release \bench{}-Core-v1 with its executable oracle, evaluation
scripts, and an evaluation card mapping each claim to its protocol.
Dataset: \url{https://huggingface.co/datasets/zhangxr7/cte-bench-core-v1}.

\end{abstract}

\section{Introduction}
\label{sec:intro}

Coding agents now change running software. They patch a service's
source, edit its stored data, or move its clock, and then keep calling
the service as if they knew how it will respond after the change. When
that expectation is wrong, the mistake rarely appears at the edited
line. It appears several calls later, as a rejected request, a wrong
balance, or a missing file. An agent that cannot anticipate these
responses cannot check its own change before relying on it.

Existing benchmarks measure neighboring abilities but not this one.
Code-execution benchmarks such as REval, CRUXEval, and code-simulation
challenges ask a model to predict what a function or snippet computes
\citep{chen2025reval,gu2024cruxeval,xu2025cruxevalx,lamalfa2024codesim};
they do not carry service state across many calls. Executable
Counterfactuals \citep{vashishtha2026ec} asks what a program would
compute under an intervention, mostly for single-call code or math
tasks. Stateful agent benchmarks
\citep{barres2025tau2,lu2025toolsandbox,pysklo2026agentdiff,
meng2026clawmark,he2026traject,huang2025stateeval}, such as
$\tau^2$-Bench, ToolSandbox, Agent-Diff, ClawMark, TRAJECT-Bench, and
StateEval, score the actions an agent chooses or the final state it
reaches. None of them fixes
the future calls in advance and asks only whether the model can predict
how a specified change alters the service's responses.

We introduce \bench{} to measure this ability directly. Its first
release, \bench{}-Core-v1, contains $255$ scenarios over six
deterministic Python services.
Figure~\ref{fig:example} shows one scenario: the source of a service,
the calls and responses observed so far, one change to the service (a
source edit or an overwrite of stored state), and $40$ future calls
fixed in advance. The model predicts the response to each future call
after the change, one call at a time. Because the future calls are
fixed, the model never chooses actions, and every prediction is checked
by running the changed service. Because we also run the same calls
without the change, we know which responses the change alters. We call
these calls \emph{effect steps} and score them separately: simply replaying the unchanged service is
correct on every other call.

We evaluate four API-hosted models under three memory protocols. With
\emph{answers revealed}, the prompt contains the recent calls with
their correct responses; with \emph{no feedback}, it contains the
observed calls and omits earlier future calls; in \emph{free rollout},
it contains the recent calls with the model's own earlier predictions.
On Core-v1, the choice of protocol changes scores far more than the
choice among these models. With answers revealed, the four models
reach $54.3\%$--$61.7\%$ effect-step value match. Without the correct
answers, all four fall to $23.2\%$--$33.2\%$. In free rollout, at most
$1.2\%$ of scenarios are predicted exactly from the first future call to
the last.

Our contributions are:
\begin{itemize}[leftmargin=*,itemsep=1pt,topsep=2pt]
\item \textbf{A benchmark for predicting the effects of changes to
stateful services.} \bench{}-Core-v1 contains $255$ scenarios over six
deterministic Python services, or $10{,}200$ predictions per model.
Every target response is produced by executing the service, so no
LLM judge is needed.
\item \textbf{A score focused on the responses the change alters.} Effect-step value
match counts exact predictions on the $2{,}476$ future calls whose
response the change alters. Replaying the unchanged service scores
$75.7\%$ on all calls but $0\%$ on effect steps.
\item \textbf{Evidence that access to correct earlier answers drives
the result.} Across four models, removing those answers lowers
effect-step value match by $28$ to $37$ points, four to five times the
$7.4$-point spread between models.
\item \textbf{An auditable release.} We release the scenarios,
executable services, graders, reference model outputs, Croissant
metadata, a datasheet, and an evaluation card that states which claims
each protocol supports.
\end{itemize}

\begin{figure}[t]
\centering
\footnotesize
\begin{minipage}[t]{0.40\linewidth}
\textbf{(a) Scenario contents}\\[3pt]
\emph{Service source}: the full edited \texttt{ratelimiter}
class. Excerpt:
\begin{lstlisting}[basicstyle=\ttfamily\scriptsize,numbers=none,frame=single,xleftmargin=0pt,aboveskip=2pt,belowskip=2pt]
41 def hit_fixed(self, key, amount=1):
42     if amount <= 1:
43         return {"error": {"code":
             "INVALID_AMOUNT", ...}}
\end{lstlisting}
\emph{Change}: a source edit applied after the last observed call; on
line 42, the token \texttt{0} has been replaced by \texttt{1}.\\[3pt]
\emph{Observed calls}, answered by the unchanged service:\\[1pt]
{\scriptsize\ttfamily
\begin{tabular}{@{}l@{\ }l@{}}
1 & advance\_clock(2) $\to$ clock 2\\
2 & hit\_sliding(k3, 1) $\to$ accepted\\
3 & reset() $\to$ cleared 1\\
4 & advance\_clock(0) $\to$ clock 2\\
5 & advance\_clock(1) $\to$ clock 3\\
\end{tabular}}\\[3pt]
\emph{Future calls}: 40 calls fixed in advance, without responses.
\end{minipage}\hfill
\begin{minipage}[t]{0.57\linewidth}
\textbf{(b) What the model must predict} (selected future calls)\\[3pt]
{\scriptsize
\setlength{\tabcolsep}{3pt}
\begin{tabular}{@{}r l l l@{}}
\toprule
\# & Future call & Without the change & After the change \\
\midrule
5  & \texttt{hit\_sliding(k2, 2)} & accepted, weight 2 & accepted, weight 2 \\
8  & \texttt{hit\_fixed(k2, 1)} & accepted, count 1 & \textbf{INVALID\_AMOUNT} $\star$ \\
9  & \texttt{hit\_fixed(k2, 3)} & accepted, count 4 & \textbf{accepted, count 3} $\star$ \\
12 & \texttt{hit\_fixed(k0, 8)} & rejected, over limit & rejected, over limit \\
30 & \texttt{hit\_fixed(k0, 1)} & accepted, count 1 & \textbf{INVALID\_AMOUNT} $\star$ \\
31 & \texttt{reset()} & cleared 2 & \textbf{cleared 1} $\star$ \\
35 & \texttt{hit\_fixed(k1, 1)} & accepted, count 1 & \textbf{INVALID\_AMOUNT} $\star$ \\
36 & \texttt{get\_count(k1, fixed)} & count 1 & \textbf{count 0} $\star$ \\
\bottomrule
\end{tabular}}\\[5pt]
\textbf{(c) How it is scored.} The six starred calls are the
\emph{effect steps}: the only future calls whose response the change
alters. Effect-step value match on this scenario is the number of
starred responses predicted exactly, divided by six. Replaying the
unchanged service answers $34$ of the $40$ future calls correctly
($85\%$ value match) but none of the six effect steps ($0\%$).
\end{minipage}
\caption{\textbf{One \bench{} scenario} (Core-v1 scenario
\texttt{eed11c9165b0}). The edit makes \texttt{hit\_fixed} reject an
amount of 1. Calls 8, 30, and 35 are rejected directly. Calls 9, 31, and
36 change only because those rejections left the counters different, so
predicting them requires carrying the changed state forward. Responses
are abbreviated here; the model predicts the full JSON response.}
\label{fig:example}
\end{figure}

\section{Related work}
\label{sec:related}

\paragraph{Runtime behavior and code simulation.} REval evaluates code
coverage prediction, program-state prediction, execution-path
prediction, output prediction, and incremental consistency on
HumanEval/ClassEval-style programs \citep{chen2025reval}. CRUXEval and
CRUXEval-X test input/output prediction for short functions, including
multilingual variants \citep{gu2024cruxeval,xu2025cruxevalx}. Code
Simulation Challenges studies line-by-line simulation of algorithmic
snippets and proposes Chain of Simulation prompting
\citep{lamalfa2024codesim}; recent reasoning-trace work analyzes
failure categories in reasoning LLMs on HumanEval+/LiveCodeBench
snippets \citep{abdollahi2025demystifying}. These benchmarks show that
predicting code execution remains difficult. \bench{} moves from
isolated functions to persistent services: the model must infer hidden
state from observed calls, apply a change, and carry the resulting state
through many future calls.

\paragraph{Counterfactual reasoning over code and structural causal models.} Executable
Counterfactuals is the closest conceptual precursor because it uses
executable code to operationalize Pearl's
abduction--intervention--prediction recipe
\citep{vashishtha2026ec}. Our interventions differ in when they act.
A source edit in \bench{} takes effect after the observed calls and
keeps the state they produced; it does not ask what would have happened
had the edited program run from the start. A state edit directly
overwrites stored state at the same point. CounterBench and broader
audits of counterfactual reasoning motivate our causal language
\citep{chen2025counterbench,yang2026ellm}, but they do not include
executable service state, source edits to a running service, or future
calls fixed in advance.

\paragraph{Stateful tool and agent evaluation.} $\tau^2$-Bench,
ToolSandbox, BFCL, NESTFUL, TRAJECT-Bench, Agent-Diff, ClawMark,
ClawsBench, and StateEval/StateGen all move beyond isolated tool calls
\citep{barres2025tau2,lu2025toolsandbox,patil2025bfcl,bfclv4,
basu2024nestful,he2026traject,pysklo2026agentdiff,meng2026clawmark,
li2026clawsbench,huang2025stateeval}. They evaluate function invocation, nested API
dependencies, action ordering, final-state correctness, or agent policy
success in a stateful environment. Interactive environments such as
WebArena, WorkArena, and OSWorld broaden ecological validity for acting
agents \citep{zhou2024webarena,drouin2024workarena,xie2024osworld}.
\bench{} is complementary: the future calls are fixed, the model does
not choose actions or change the environment, and the score is the
accuracy of its predicted responses after a specified change.

\paragraph{Code world models and executable environments.} Code-world
model work studies whether models can predict program state during
execution. \citet{rahmani2026debugcwm} identifies token-budget exhaustion,
string-valued state, and action-generation errors as common failure
sources, and \citet{guo2025sample} train state prediction as an
internal environment model for stateful tool use. Agent
World Model and Agent-World scale the synthesis of executable,
database-backed, or tool-rich environments
\citep{wang2026awm,dong2026agentworld}. These systems are promising
sources of future \bench{} services. \bench{} is narrower: it uses deterministic services with specified
source and state edits, and grades predicted responses by executing
the services rather than by judging an agent's actions. Appendix~\ref{app:axis-delta} gives a
feature-by-feature comparison.

\section{The \bench{} benchmark}
\label{sec:formulation}

\subsection{Task}
\label{sec:task}

A \emph{scenario} is one prediction task with four parts, all shown in
Figure~\ref{fig:example}: the source $P$ of a Python service; the
\emph{observed calls} $\tau$, a sequence of calls and the responses the
unchanged service gave to them; an \emph{intervention} $\xi$ applied
after the last observed call; and a list $Q$ of $40$ \emph{future
calls}. The target responses are produced by the service itself, which
we call the \emph{oracle}: a fresh instance replays $\tau$, applies
$\xi$, and answers each call in $Q$. The model predicts the $40$ responses with one request per future
call. Each request contains $P$, $\xi$, the call to predict, and the
earlier calls specified by the memory protocol
(Section~\ref{sec:protocols}). For source edits the prompt shows the
edited source together with the line and tokens that were changed.

The oracle also answers $Q$ without applying $\xi$. We call these
responses the \emph{no-intervention trace} and the responses after
$\xi$ the \emph{target trace}. An \emph{effect step} is a future call
whose response differs between the two traces. Each released scenario
stores both traces and the final service states, so every score can be
recomputed without querying a model or an LLM judge.

\subsection{Interventions}
\label{sec:interventions}

A \textbf{source edit} changes one token in the service code, such as
inverting a comparison or changing a numeric threshold. The edit takes
effect after the observed calls: the state they built is kept, and only
the future calls run the edited code. The question is therefore what
happens from this point on, not what would have happened had the edited
code run from the start. Source edits are admitted only when the edited
service can load the existing state and replay deterministically.

A \textbf{state edit} leaves the code unchanged and overwrites stored
state at the same point. In Core-v1, state edits either move the
service clock or overwrite a stored rate-limit counter. They model
external changes to stored service state; we do not claim that they are
the most common change a developer makes. In both cases
the edit is small, but its effects depend on the state built by the
observed calls.

\subsection{Services and dataset}
\label{sec:dataset}

The six deterministic services cover common backend state machines and
compact ports of open-source algorithms. \texttt{auth} issues and
checks tokens with expiry times, \texttt{bank} keeps a ledger with
overdraft checks, \texttt{cart} tracks shopping-cart quantities and
checkout, \texttt{filesystem} is an in-memory file system,
\texttt{ratelimiter} implements fixed- and sliding-window rate limits,
and \texttt{reservation} enforces booking quotas.

\bench{}-Core-v1 contains $255$ scenarios in nine service--intervention
cells (Appendix~\ref{app:per-kind}). Of these, $210$ use source edits
($150$ branch inversions and $60$ threshold changes, drawn from $49$
distinct patches) and $45$ use state edits ($30$ clock jumps and $15$
counter overwrites). Each service contributes $30$ scenarios except
\texttt{ratelimiter}, which contributes $105$ across four cells.
Observed histories contain $5$ to $56$ calls. Every scenario is scored
on its first $40$ future calls, giving $255 \times 40 = 10{,}200$
predictions per model. The $40$ predictions within a scenario are not
independent, so all confidence intervals resample scenarios.

We keep only scenarios in which the intervention changes at least one
response in the stored trace, which extends beyond the $40$ scored
calls. In $8$ scenarios the first change comes after call $40$, so
these contribute no effect steps (Appendix~\ref{app:filter-v2}).
Core-v1 therefore tests
whether a model predicts changes that do happen. The complementary
error, inventing a change that does not happen, is scored on no-effect
companions: the release pairs every Core-v1 scenario with a companion
that shares its service, history, and future calls, and results are
reported as the pair (effect-step VM, no-effect VM)
(Appendix~\ref{app:no-effect-v1}).

\subsection{Scores}
\label{sec:metrics}

\textbf{Value match (VM)} counts a prediction as correct only if the
parsed JSON response equals the oracle response exactly.
\textbf{Effect-step VM}, our main score, is VM restricted to effect
steps; Core-v1 has $2{,}476$ of them. Section~\ref{sec:exp-baselines}
shows why this restriction is needed. \textbf{Status match (SM)} checks
only whether the prediction and the oracle agree on success versus
error, and serves as a diagnostic. \textbf{Field match} flattens each
oracle response into its scalar fields and gives partial credit for
each field predicted correctly; we use it to find near misses that
exact VM hides. Calls that return a provider error and no model
output are not scored; the release lists the count for every run. The
appendix also reports a source-diff localization diagnostic, which
does not measure response prediction, and explains why an early state
probe is excluded from Core-v1
(Appendices~\ref{app:expl-prompt}--\ref{app:sp-v2}).

\subsection{Memory protocols}
\label{sec:protocols}

Every prompt contains the service source, the intervention, and the
future call to predict. The three protocols differ in which earlier
calls the prompt includes.
\begin{itemize}[leftmargin=*,itemsep=1pt,topsep=2pt]
\item \textbf{Answers revealed.} The prompt includes the most recent
$20$ calls, observed or future, and each earlier future call appears
with its correct response. Earlier prediction errors never enter later
prompts.
\item \textbf{No feedback.} The prompt includes all observed calls and
omits earlier future calls. The model must predict each response from
the observed history and the intervention alone.
\item \textbf{Free rollout.} The prompt uses the same $20$-call window
as answers revealed, but each earlier future call appears with the
model's own predicted response. Early errors can therefore carry
forward.
\end{itemize}
The protocols answer different questions. Answers revealed asks
whether the model can predict the next response when the recent history
is correct. No feedback measures how well the model predicts each response from
the observed history and the intervention alone. Free rollout is closest to an
agent relying on its own expectations, because the model must live with
its earlier predictions. We report the protocol next to every score.

\section{Results}
\label{sec:experiments}

\paragraph{Setup.} We evaluate four API-hosted model configurations:
DeepSeek~V4-Flash, Kimi~K2.5, Qwen3.6-35B-A3B, and Claude~Sonnet~4.6
\citep{deepseek2026v4,moonshot2026k25,qwen2026qwen36a3b,
anthropic2026adaptive}. The first three have openly released weights;
Sonnet~4.6 is proprietary and runs with the provider's default reasoning
setting, so it is a reference configuration rather than a controlled
comparison point. Each future call is one request with a strict
JSON output format. We report means with $95\%$ bootstrap intervals
that resample the $255$ scenarios ($B{=}2000$). Serving routes,
reasoning settings, and cost are listed in
Appendix~\ref{app:model-config}.

\subsection{Why score effect steps?}
\label{sec:exp-baselines}

Table~\ref{tab:baselines} compares the models with three predictors
that use no model. Replaying the service without the intervention
matches $75.7\%$ of all future responses, higher than every model's
all-call VM, because most future calls are unaffected by the change.
By construction, it matches none of the effect steps. Copying the last
observed response or predicting the most common observed status scores
near zero VM but $72.3\%$ and $76.8\%$ SM, so predicting only success or
error is easy. All-call VM thus rewards ignoring the change; we use
effect-step VM as the main score and SM only as a diagnostic.

\begin{table}[H]
\centering\footnotesize
\caption{\textbf{All-call VM rewards ignoring the change.} Predictors
without a model versus the four models (answers revealed), in \%.
Eff.\ columns are restricted to the $2{,}476$ effect steps, excluding
calls that returned a provider error.}
\label{tab:baselines}
\begin{tabular}{l|rrrr}
\toprule
Predictor & VM & Eff.\,VM & SM & Eff.\,SM \\
\midrule
Replay without the intervention & 75.7 & 0.0 & 91.6 & 65.5 \\
Copy the last observed response & 1.5 & 1.1 & 72.3 & 60.5 \\
Most common observed status & 0.0 & 0.0 & 76.8 & 60.5 \\
\midrule
DeepSeek V4-Flash & 75.1 & 61.2 & 92.8 & 86.7 \\
Kimi K2.5 & 72.0 & 61.7 & 90.0 & 86.0 \\
Qwen3.6-35B-A3B & 65.6 & 54.3 & 84.8 & 79.4 \\
Claude Sonnet 4.6 & 61.8 & 58.4 & 80.7 & 75.7 \\
\bottomrule
\end{tabular}

\end{table}

\subsection{How well do models predict changed responses?}
\label{sec:exp-memory}

\begin{figure}[t]
\centering
\includegraphics[width=0.72\linewidth]{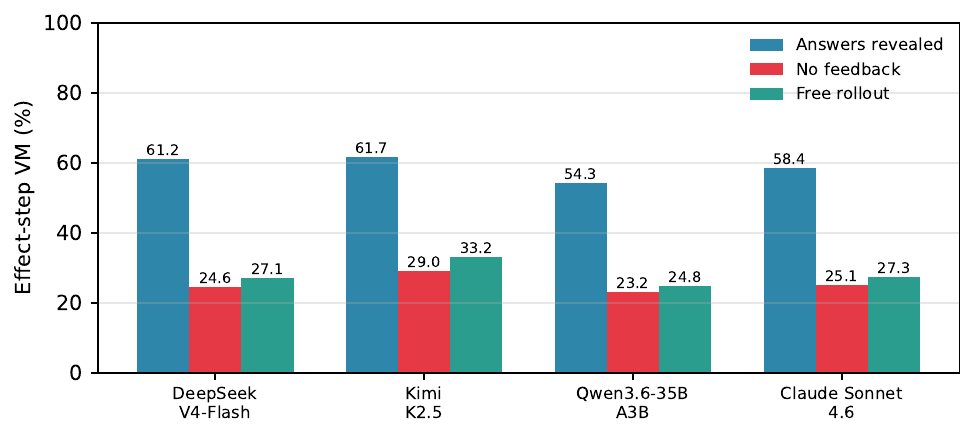}
\caption{\textbf{The protocol changes scores more than the choice
among these models.} Effect-step VM on Core-v1 under the three memory
protocols (Section~\ref{sec:protocols}).}
\label{fig:hero}
\end{figure}

Figure~\ref{fig:hero} and Table~\ref{tab:main-results} give the main
scorecard. With answers revealed, DeepSeek, Kimi, and Sonnet reach
$58.4\%$--$61.7\%$ effect-step VM with overlapping intervals. Qwen has
the lowest point estimate, $54.3\%$, and its interval overlaps
Sonnet's. Removing the correct earlier answers lowers every model by
$28$ to $37$ points, to $23.2\%$--$29.0\%$ with no feedback and
$24.8\%$--$33.2\%$ in free rollout. This drop is four to five times the
$7.4$-point spread between models under answers revealed.

Each protocol also has a reference that performs no simulation: repeat
the most recent visible response to the same call. It reaches $19.3\%$
effect-step VM with answers revealed, $3.3\%$ with no feedback, and
$2.2\%$ in free rollout (Table~\ref{tab:main-results};
Appendix~\ref{app:reference-policies}). Revealed answers therefore help
even a policy that only copies, but every model stays $35$ to $42$
points above this reference with answers revealed and $20$ to $31$
points above it under the other two protocols.

Parse failures do not explain the drop. At most $0.5\%$ of the
responses from DeepSeek, Kimi, and Qwen are unparseable under any
protocol; with no feedback, $3$ of DeepSeek's $10{,}200$ responses are
unparseable (Appendix~\ref{app:error-modes}). For Sonnet, $13.5\%$ to
$16.7\%$ of responses are unparseable under every protocol, including
$14.3\%$ with answers revealed, so formatting lowers all of its scores but does not
account for the gap between protocols.

A \bench{} score is thus meaningful only together with its protocol. With
answers revealed, it measures one-step prediction from a correct recent
history, not the ability to simulate the changed service unaided.

\begin{table}[t]
\centering\small
\caption{\textbf{Core-v1 scorecard} ($255$ scenarios, \%): mean $\pm$
half-width of the $95\%$ bootstrap interval over scenarios. VM is exact
response match and SM is success/error status match. Eff.\,VM uses the
$2{,}476$ effect steps and VM and SM the $10{,}200$ future calls,
excluding calls that returned a provider error (Section~\ref{sec:metrics}).
$^{\dagger}$Repeats the most recent visible response to the same call
(operation and arguments) and performs no simulation.}
\label{tab:main-results}
\begin{tabular}{l|rrr}
\toprule
Model & Eff.\,VM & VM & SM \\
\midrule
\multicolumn{4}{l}{\emph{Answers revealed}} \\
DeepSeek V4-Flash & 61.2\,{\scriptsize$\pm$3.1} & 75.1\,{\scriptsize$\pm$1.4} & 92.8\,{\scriptsize$\pm$1.0} \\
Kimi K2.5 & 61.7\,{\scriptsize$\pm$3.1} & 72.0\,{\scriptsize$\pm$1.4} & 90.0\,{\scriptsize$\pm$1.0} \\
Qwen3.6-35B-A3B & 54.3\,{\scriptsize$\pm$2.8} & 65.6\,{\scriptsize$\pm$1.4} & 84.8\,{\scriptsize$\pm$1.1} \\
Claude Sonnet 4.6 & 58.4\,{\scriptsize$\pm$3.7} & 61.8\,{\scriptsize$\pm$2.3} & 80.7\,{\scriptsize$\pm$1.4} \\
\emph{Copy-last reference}$^{\dagger}$ & 19.3\,{\scriptsize$\pm$2.3} & 18.5\,{\scriptsize$\pm$1.3} & 80.2\,{\scriptsize$\pm$1.8} \\
\midrule
\multicolumn{4}{l}{\emph{No feedback}} \\
DeepSeek V4-Flash & 24.6\,{\scriptsize$\pm$3.5} & 39.8\,{\scriptsize$\pm$1.9} & 85.5\,{\scriptsize$\pm$1.7} \\
Kimi K2.5 & 29.0\,{\scriptsize$\pm$3.8} & 38.6\,{\scriptsize$\pm$1.8} & 84.1\,{\scriptsize$\pm$1.6} \\
Qwen3.6-35B-A3B & 23.2\,{\scriptsize$\pm$3.2} & 34.9\,{\scriptsize$\pm$1.7} & 81.0\,{\scriptsize$\pm$1.6} \\
Claude Sonnet 4.6 & 25.1\,{\scriptsize$\pm$3.5} & 29.5\,{\scriptsize$\pm$2.0} & 74.9\,{\scriptsize$\pm$1.9} \\
\emph{Copy-last reference}$^{\dagger}$ & 3.3\,{\scriptsize$\pm$1.2} & 9.3\,{\scriptsize$\pm$1.3} & 78.0\,{\scriptsize$\pm$2.0} \\
\midrule
\multicolumn{4}{l}{\emph{Free rollout}} \\
DeepSeek V4-Flash & 27.1\,{\scriptsize$\pm$4.0} & 57.5\,{\scriptsize$\pm$2.4} & 88.0\,{\scriptsize$\pm$1.6} \\
Kimi K2.5 & 33.2\,{\scriptsize$\pm$3.7} & 52.3\,{\scriptsize$\pm$2.1} & 83.9\,{\scriptsize$\pm$1.6} \\
Qwen3.6-35B-A3B & 24.8\,{\scriptsize$\pm$3.4} & 43.8\,{\scriptsize$\pm$2.1} & 74.8\,{\scriptsize$\pm$2.0} \\
Claude Sonnet 4.6 & 27.3\,{\scriptsize$\pm$4.0} & 29.6\,{\scriptsize$\pm$2.6} & 69.3\,{\scriptsize$\pm$2.5} \\
\emph{Copy-last reference}$^{\dagger}$ & 2.2\,{\scriptsize$\pm$1.0} & 4.0\,{\scriptsize$\pm$0.7} & 77.6\,{\scriptsize$\pm$2.0} \\
\bottomrule
\end{tabular}

\end{table}

\subsection{How do errors build up in free rollout?}
\label{sec:exp-rollout}

\begin{figure}[H]
\centering
\includegraphics[width=0.62\linewidth]{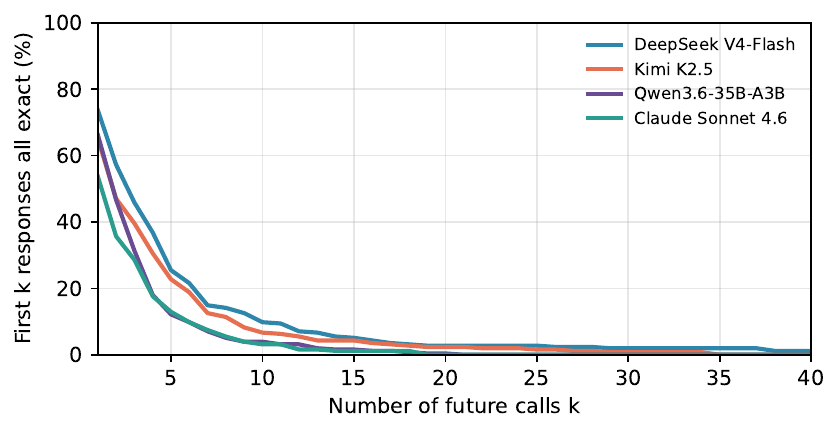}
\caption{\textbf{Share of scenarios still exact after $k$ future
calls} in free rollout. Values at selected $k$ are in
Appendix~\ref{app:rollout-survival}.}
\label{fig:rollout-survival}
\end{figure}

In free rollout the model conditions on its own earlier predictions,
so one wrong response can corrupt later ones.
Table~\ref{tab:rollout-summary} and Figure~\ref{fig:rollout-survival}
show that this happens early. In the median scenario, the first wrong
response is at future call $2$ or $3$. At most $9.8\%$ of scenarios are
still exact after $10$ calls, and only DeepSeek predicts any scenario
exactly over all $40$ calls ($1.2\%$). VM also falls from calls 1--10
to calls 21--40 for every model. Whole-trace exact match therefore barely separates the models, so we
also report call-level VM and the survival curve.

\begin{table}[H]
\centering\footnotesize
\caption{\textbf{Free-rollout errors start early and accumulate.}
Whole-trace exact is the share of scenarios (\%) with all $40$
responses exact. The first wrong call is the first future call, counted
from $1$, whose prediction is not an exact match; an unparseable response
counts as a wrong prediction. The $3$ DeepSeek scenarios without a wrong call are excluded
from its mean and median only. The last two columns give all-call VM (\%) over future calls
1--10 and 21--40.}
\label{tab:rollout-summary}
\begin{tabular}{l|rrrrr}
\toprule
& Whole trace & \multicolumn{2}{c}{First wrong call} & \multicolumn{2}{c}{VM by call range} \\
Model & exact & mean & median & 1--10 & 21--40 \\
\midrule
DeepSeek V4-Flash & 1.2 & 4.6 & 3 & 66.9 & 53.6 \\
Kimi K2.5 & 0.0 & 4.2 & 2 & 61.3 & 47.5 \\
Qwen3.6-35B-A3B & 0.0 & 3.2 & 2 & 53.3 & 39.2 \\
Claude Sonnet 4.6 & 0.0 & 2.9 & 2 & 40.4 & 24.4 \\
\bottomrule
\end{tabular}

\end{table}

\subsection{Where do errors concentrate?}
\label{sec:exp-perdomain}

\begin{figure}[H]
\centering
\includegraphics[width=0.82\linewidth]{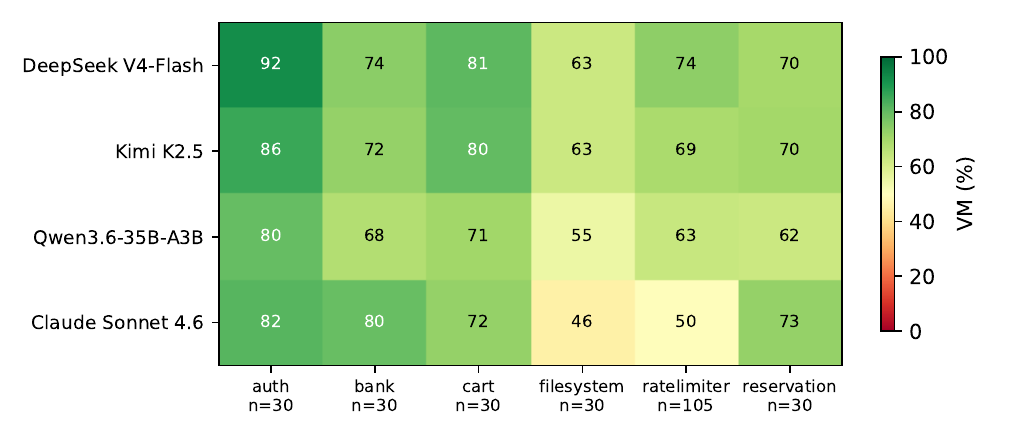}
\caption{\textbf{Failures are service-specific.} All-call VM (\%) per
model and service under answers revealed; $n$ is the number of
scenarios. This is a descriptive profile and is not used for pairwise
significance claims.}
\label{fig:heatmap}
\end{figure}

Table~\ref{tab:effect-non-effect-field} and Figure~\ref{fig:heatmap}
break the answers-revealed results down in two ways. First, unchanged
calls are not trivial: VM on non-effect calls ranges from $62.9\%$ to
$79.6\%$, higher than on effect steps but far below the $100\%$ that
replay would achieve. Second, exact matching hides some near misses:
field match on effect steps exceeds effect-step VM by $9.1$--$11.4$
points for DeepSeek, Kimi, and Qwen, but by only $0.5$ points for
Sonnet.

By service, Sonnet has the highest VM of the four models on
\texttt{bank} and \texttt{reservation} and the lowest on
\texttt{filesystem} and \texttt{ratelimiter}, where DeepSeek and Kimi
score higher. These profiles are descriptive, not rankings: five
services have $30$ scenarios each and \texttt{ratelimiter} has $105$,
so intervals are wide (Appendix~\ref{app:per-domain}), and
weighting services equally changes the averages
(Appendix~\ref{app:domain-weight}).

\begin{table}[H]
\centering\footnotesize
\caption{\textbf{Unchanged calls are not trivial, and exact match hides
near misses} (answers revealed, \%). Non-effect VM covers future calls
whose response the change does not alter. Field match gives partial
credit for each correctly predicted scalar field of the JSON response;
Eff.\ field match covers effect steps and field match covers all
future calls.}
\label{tab:effect-non-effect-field}
\begin{tabular}{l|rrrr}
\toprule
Model & Eff.\,VM & Non-effect VM & Eff.\,field match & Field match \\
\midrule
DeepSeek V4-Flash & 61.2 & 79.6 & 72.6 & 82.8 \\
Kimi K2.5 & 61.7 & 75.3 & 71.3 & 78.2 \\
Qwen3.6-35B-A3B & 54.3 & 69.2 & 63.4 & 73.7 \\
Claude Sonnet 4.6 & 58.4 & 62.9 & 58.9 & 62.2 \\
\bottomrule
\end{tabular}
\end{table}

\paragraph{No-effect scenarios.} Effect-step VM cannot penalize a model
for predicting a change that does not happen. On $106$ scenarios whose
intervention changes no future response, the models reach only
$82.3\%$--$84.5\%$ VM with answers revealed, although replaying the
unchanged service would be exact (Appendix~\ref{app:no-effect}). The
release therefore pairs every Core-v1 scenario with a no-effect
companion on the same service, history, and future calls;
Appendix~\ref{app:no-effect-v1} reports the paired scores for
DeepSeek~V4-Flash and Kimi~K2.5.

\paragraph{Reasoning and renaming.} Turning on reasoning raises
DeepSeek~V4-Flash from $24.6\%$ to $34.5\%$ effect-step VM with no
feedback and from $62.3\%$ to $81.1\%$ with answers revealed ($233$
scenarios; Appendix~\ref{app:reasoning-full}), so the gap between
protocols stays large with reasoning. An identifier-renamed copy of the
$210$ source-edit scenarios tests how much models rely on familiar
names; renaming lowers Kimi~K2.5's effect-step VM by $2.7$ points
(Appendix~\ref{app:renamed}).

\section{Discussion and limitations}
\label{sec:discussion}

\paragraph{What the protocol gap means.} To answer a future call, a
model must infer the service state from the observed calls, apply the
intervention, simulate the changed service, and stay consistent with
its earlier answers. Revealing the correct earlier answers does much of
the state tracking for the model. The $28$- to $36$-point drop without
them shows that current models depend heavily on this help. For a coding
agent that relies on its own expectation of a changed service, the
free-rollout score is the relevant one, and it is low.

\paragraph{Status versus payload.} SM stays between $69.3\%$ and
$92.8\%$ under every protocol while VM falls sharply: models predict whether a call succeeds more
often than what it returns. A score based only on success or error
would miss these errors, which matter when a
debugging decision depends on a balance, a counter, a path, or an
expiry time. Exact VM is strict, but every oracle response is produced
by the same serializer, so exact equality is well defined; field match
shows how much partial credit this strictness removes.

\paragraph{What the score does not isolate.} Effect-step VM is an
end-to-end score. A wrong response can come from a wrong inferred
state, a misapplied intervention, a simulation error, or a formatting
error, and Core-v1 does not attribute errors to these steps. The
per-service and per-intervention views are descriptive failure
profiles, not ablations. Because each scenario stores the oracle
states, targeted follow-ups are straightforward: revealing the true
state at the intervention point would separate state inference from
simulation, and allowing code execution would measure how far
text-only prediction is from running the service.

\paragraph{Choosing a protocol.} Use answers revealed to compare
one-step prediction when the recent history is correct. Use free
rollout to ask whether a model can simulate a changed service on its
own. Use no feedback to measure prediction from the observed history and
the intervention alone. Report the protocol with
every score; Appendix~\ref{app:eval-card} lists the reporting details
needed for comparable results.

\paragraph{Limitations.} Core-v1 covers six deterministic Python
services, and \texttt{ratelimiter} is over-represented; we report an
equal-service average as a sensitivity check. The interventions are
synthetic source edits and state overwrites, and models predict
responses from text alone. We evaluate four API-hosted configurations
with one prompt format. Sensitivity to prompt wording is unmeasured,
and the results support no claims about model scale or training
objective. The controlled reasoning comparison covers one model,
DeepSeek~V4-Flash. Core-v1
contains only scenarios in which the intervention changes some
response; the no-effect companions (Appendix~\ref{app:no-effect-v1})
score the complementary error of predicting a change that does not
happen. Finally, \bench{} is not a production safety certification: a
high score does not show that an agent's changes are safe to deploy.

\paragraph{Intended use and dual use.} \bench{} is meant for controlled
failure profiling and robustness evaluation, not for vulnerability
discovery, attack planning, or production security certification. Its
services are small deterministic testbeds with no production targets or
real credentials, and the model only predicts responses to fixed calls:
it chooses no actions, calls no tools, and never changes the
environment. These constraints limit direct misuse but do not prevent
the underlying capability from being adapted elsewhere, and we do not
evaluate transfer to offensive workflows.

\section{Conclusion}
\label{sec:conclusion}

\bench{} asks whether a model can predict how a specified change alters
the future responses of a stateful service, and checks every prediction
by running the service. On Core-v1, four current models reach $54.3\%$--$61.7\%$ effect-step
VM when the correct earlier responses are supplied. Without them, they
fall to $23.2\%$--$33.2\%$,
and in free rollout their first error typically comes within the first
three calls. We release the scenarios, executable services, graders,
reference outputs, and an evaluation card that states which claims each
protocol supports, so that new models can be compared on the same
terms.

\clearpage
\bibliographystyle{plainnat}
\bibliography{references}

\appendix
\section{Feature-by-feature comparison with related benchmarks}
\label{app:axis-delta}

Table~\ref{tab:axis-delta} expands the comparison in
Sections~\ref{sec:intro} and~\ref{sec:related}.

\begin{table}[ht]
\centering\small
\caption{\bench{} versus adjacent benchmark families. ``Source edit
after history'' is a code edit that takes effect after the observed
calls; ``state overwrite'' directly changes stored state at the same
point. Agent benchmarks score the actions a policy chooses, not
predicted responses to fixed calls.}
\label{tab:axis-delta}
\resizebox{\linewidth}{!}{%
\begin{tabular}{l|ccccc}
\toprule
Feature & EC & $\tau^2$/Agent-Diff/ClawMark & TRAJECT-Bench & Debug-CWMs & \textbf{\bench{}} \\
\midrule
Persistent state          & no  & yes & yes & yes & \textbf{yes} \\
Observed history          & 1 obs. & agent rollout & tool call & varies & \textbf{5--56 calls} \\
Source edit after history & no  & no  & no  & no  & \textbf{yes} \\
State overwrite           & no  & no  & no  & no  & \textbf{yes} \\
Target                    & 1 scalar & final state & tool-call fidelity & internal probes & \textbf{40 future responses} \\
Oracle                    & modulo-$n$ & scripted user & tool dispatch & hand-written & \textbf{executed service} \\
Evaluates                 & reasoning & agent policy & agent policy & model internals & \textbf{response prediction} \\
\bottomrule
\end{tabular}}
\end{table}

\section{Model configurations and cost}
\label{app:model-config}
\label{app:billing}

Table~\ref{tab:model-scope} lists the evaluated configurations.
DeepSeek~V4-Flash is served through the \texttt{deepseek-chat}
compatibility alias; we list serving route and reasoning control
separately so that API aliases are not mistaken for distinct
checkpoints. We use APIs for uniform logging, retries, and cost
accounting. Parameter counts are reported for context only.

\begin{table}[ht]
\centering\footnotesize
\caption{Evaluated model configurations. Sonnet~4.6 runs with the
provider-default reasoning setting and is a reference configuration.}
\label{tab:model-scope}
\begin{tabularx}{\linewidth}{l l l X}
\toprule
Reported configuration & Serving route & Reported size & Reasoning control \\
\midrule
DeepSeek V4-Flash / chat alias & DeepSeek API & 284B total / 13B active MoE & non-thinking API mode \\
Kimi K2.5 & Bedrock/Moonshot & 1.1T-parameter MoE & thinking disabled \\
Qwen3.6-35B-A3B & OpenRouter-routed & 35B total / 3B active MoE & reasoning requested off; usage audited \\
Claude Sonnet 4.6 & Amazon Bedrock & undisclosed proprietary & thinking not explicitly requested; provider default \\
\bottomrule
\end{tabularx}

\end{table}

Oracle generation, grading, and table construction are CPU-only Python
workloads; the benchmark itself uses no GPU training. Model evaluation
is dominated by hosted API inference and can be parallelized over
scenarios. Cost depends strongly on provider caching, output length,
and provider choice. At the prices in effect during collection, the
additional free-rollout runs cost \$1.30 for DeepSeek~V4-Flash and \$12.59 for
Kimi~K2.5. These are the costs of those two runs, not a uniform
per-model price; a full Core-v1 run is $10{,}200$ predictions per model
and protocol. We report API billing; the providers expose no hardware,
energy, or carbon telemetry, so billing and prediction counts do not
support a carbon estimate. Prompt and pipeline checks can use a few
services or scenarios, whereas comparable scores require the full
split. The scenarios, oracle, and grader are model-independent, and any
hosted or locally served model can be evaluated; three of the four
evaluated model families have openly released weights. Reproducing a
reported row exactly requires the same model version and serving route,
so the evaluation card (Appendix~\ref{app:eval-card}) asks every report
to state them together with the inference cost.

\section{Reference policies}
\label{app:reference-policies}

Table~\ref{tab:reference-policies} scores policies that receive exactly
the information a model receives under each protocol but perform no
simulation. Replaying the service without the intervention predicts the
unchanged response. The copy policies repeat the most recent visible
response to the same operation, or to the same call (operation and
arguments), and predict an empty object when there is none. Visible
responses are exactly those in the prompt's call history: with answers
revealed and in free rollout, the most recent $20$ calls (observed calls
followed by future calls, with their correct responses or the policy's
own earlier predictions); with no feedback, all observed calls. A model's margin
over the reference for its protocol is a descriptive calibration point,
not a decomposition of the model's score.
\texttt{scripts/reference\_policies.py} computes the table from the
released scenarios with the released grader.

\begin{table}[ht]
\centering\footnotesize
\caption{Non-simulating reference policies on Core-v1 (\%), with
half-widths of $95\%$ bootstrap intervals over scenarios.}
\label{tab:reference-policies}
\begin{tabular}{l|rrr}
\toprule
Reference policy & Eff.\,VM & VM & SM \\
\midrule
\multicolumn{4}{l}{\emph{Answers revealed}} \\
Replay without the intervention & 0.0\,{\scriptsize$\pm$0.0} & 75.7\,{\scriptsize$\pm$2.3} & 91.6\,{\scriptsize$\pm$1.1} \\
Copy last response, same operation & 14.3\,{\scriptsize$\pm$2.2} & 11.5\,{\scriptsize$\pm$0.9} & 83.7\,{\scriptsize$\pm$1.2} \\
Copy last response, same call & 19.3\,{\scriptsize$\pm$2.3} & 18.5\,{\scriptsize$\pm$1.3} & 80.2\,{\scriptsize$\pm$1.8} \\
\midrule
\multicolumn{4}{l}{\emph{No feedback}} \\
Copy last response, same operation & 1.8\,{\scriptsize$\pm$0.8} & 3.5\,{\scriptsize$\pm$0.5} & 77.9\,{\scriptsize$\pm$2.0} \\
Copy last response, same call & 3.3\,{\scriptsize$\pm$1.2} & 9.3\,{\scriptsize$\pm$1.3} & 78.0\,{\scriptsize$\pm$2.0} \\
\midrule
\multicolumn{4}{l}{\emph{Free rollout}} \\
Copy last response, same operation & 1.4\,{\scriptsize$\pm$0.7} & 3.1\,{\scriptsize$\pm$0.5} & 77.3\,{\scriptsize$\pm$2.0} \\
Copy last response, same call & 2.2\,{\scriptsize$\pm$1.0} & 4.0\,{\scriptsize$\pm$0.7} & 77.6\,{\scriptsize$\pm$2.0} \\
\bottomrule
\end{tabular}

\end{table}

\section{Per-intervention value match}
\label{app:per-kind}

Table~\ref{tab:per-kind} gives Core-v1 all-call VM by intervention
type under answers revealed.
\begin{table}[ht]
\centering\footnotesize
\caption{Core-v1 all-call VM (\%) by intervention type, answers
revealed. The delayed-effect patches of
Appendix~\ref{app:reasoning-probe} are a separate probe.}
\label{tab:per-kind}
\begin{tabular}{l|rrrr}
\toprule
Model & Branch inversion & Clock jump & Counter overwrite & Threshold change \\
\midrule
DeepSeek V4-Flash & 73.0 & 68.7 & 70.5 & 84.8 \\
Kimi K2.5 & 71.6 & 64.6 & 68.8 & 77.4 \\
Qwen3.6-35B-A3B & 64.2 & 60.7 & 60.5 & 72.7 \\
Claude Sonnet 4.6 & 63.5 & 48.7 & 48.2 & 67.6 \\
\bottomrule
\end{tabular}
\end{table}

\section{Per-service value match}
\label{app:per-domain}

Table~\ref{tab:per-domain} gives Core-v1 all-call VM per service with
bootstrap intervals over scenarios. These are descriptive and are not
pairwise significance tests.

\begin{table}[ht]
\centering\footnotesize
\caption{Core-v1 all-call VM (\%) per service under answers revealed,
with scenario counts in parentheses and half-widths of $95\%$
bootstrap intervals.}
\label{tab:per-domain}
\begin{tabular}{l|rrrrrr}
\toprule
Model & auth (30) & bank (30) & cart (30) & filesystem (30) & ratelimiter (105) & reservation (30) \\
\midrule
DeepSeek V4-Flash & 91.9\,{\scriptsize$\pm$1.8} & 73.9\,{\scriptsize$\pm$2.7} & 81.2\,{\scriptsize$\pm$3.7} & 62.8\,{\scriptsize$\pm$3.7} & 73.8\,{\scriptsize$\pm$1.8} & 69.9\,{\scriptsize$\pm$3.4} \\
Kimi K2.5 & 85.6\,{\scriptsize$\pm$1.7} & 72.3\,{\scriptsize$\pm$2.5} & 80.2\,{\scriptsize$\pm$3.6} & 62.7\,{\scriptsize$\pm$2.7} & 68.8\,{\scriptsize$\pm$1.8} & 70.2\,{\scriptsize$\pm$3.7} \\
Qwen3.6-35B-A3B & 79.9\,{\scriptsize$\pm$3.4} & 67.5\,{\scriptsize$\pm$3.0} & 70.7\,{\scriptsize$\pm$3.7} & 55.0\,{\scriptsize$\pm$3.1} & 63.3\,{\scriptsize$\pm$1.7} & 62.5\,{\scriptsize$\pm$3.5} \\
Claude Sonnet 4.6 & 81.8\,{\scriptsize$\pm$2.8} & 79.6\,{\scriptsize$\pm$2.6} & 72.4\,{\scriptsize$\pm$6.0} & 45.6\,{\scriptsize$\pm$5.1} & 49.6\,{\scriptsize$\pm$2.1} & 72.6\,{\scriptsize$\pm$3.5} \\
\bottomrule
\end{tabular}
\end{table}

\section{Equal-service weighting}
\label{app:domain-weight}

Core-v1 weights scenarios equally. Because \texttt{ratelimiter}
contributes four intervention cells and therefore more scenarios than
the other services, Table~\ref{tab:domain-weight} also reports an
average that weights the six services equally. This is a sensitivity
check, not a replacement scorecard.

\begin{table}[ht]
\centering\footnotesize
\caption{Equal-service weighting under answers revealed (\%).
``Service-macro'' averages the six per-service rates equally; the other
columns weight scenarios equally.}
\label{tab:domain-weight}
\begin{tabular}{l|rrrr}
\toprule
Model & Eff.\,VM & Service-macro Eff.\,VM & VM & Service-macro VM \\
\midrule
DeepSeek V4-Flash & 61.2 & 64.7 & 75.1 & 75.6 \\
Kimi K2.5 & 61.7 & 65.9 & 72.0 & 73.3 \\
Qwen3.6-35B-A3B & 54.3 & 59.1 & 65.6 & 66.5 \\
Claude Sonnet 4.6 & 58.4 & 61.7 & 61.8 & 66.9 \\
\bottomrule
\end{tabular}
\end{table}

\section{Reasoning off and on for DeepSeek V4-Flash}
\label{app:reasoning-full}

DeepSeek~V4-Flash exposes a reasoning-off mode (\texttt{deepseek-chat},
the configuration of Table~\ref{tab:main-results}) and a reasoning-on
mode (\texttt{deepseek-reasoner}) through the same API, so it supports a
controlled comparison on Core-v1. The reasoning runs use reasoning
effort \texttt{high} and a $12{,}000$-token output budget; all other
settings equal those of Table~\ref{tab:main-results}.
Table~\ref{tab:reasoning-full} covers all $255$ scenarios with no
feedback and the $233$ scenarios of the reasoning run with answers
revealed, with the reasoning-off row computed on the same scenarios.
Reasoning raises effect-step VM by $9.9$ points with no feedback and by
$18.8$ points with answers revealed and lowers the share of parseable
responses. With reasoning, no-feedback effect-step VM ($34.5\%$) stays far
below answers-revealed effect-step VM without reasoning ($61.2\%$ on all
$255$ scenarios, Table~\ref{tab:main-results}).
\texttt{scripts/score\_reasoning\_pairs.py} reproduces the table from the
released outputs.

\begin{table}[ht]
\centering\footnotesize
\caption{Reasoning off and on for DeepSeek~V4-Flash on Core-v1 (\%).
Eff.\,VM uses the $2{,}296$ (answers revealed) and $2{,}476$ (no feedback)
effect steps of the listed scenarios, excluding calls that returned a
provider error; $\pm$ is the half-width of a
$95\%$ bootstrap interval over scenarios; WF is the share of parseable
responses. Reasoning-off runs: April 2026; reasoning-on runs: July 2026.}
\label{tab:reasoning-full}
\begin{tabular}{l|l|rrrr}
\toprule
Protocol (scenarios) & Reasoning & Eff.\,VM & VM & SM & WF \\
\midrule
Answers revealed (233) & off & 62.3\,{\scriptsize$\pm$3.0} & 75.3\,{\scriptsize$\pm$1.5} & 92.3\,{\scriptsize$\pm$1.1} & 100.0 \\
 & on & 81.1\,{\scriptsize$\pm$2.2} & 84.4\,{\scriptsize$\pm$1.3} & 91.4\,{\scriptsize$\pm$0.9} & 94.5 \\
\midrule
No feedback (255) & off & 24.6\,{\scriptsize$\pm$3.5} & 39.8\,{\scriptsize$\pm$1.9} & 85.5\,{\scriptsize$\pm$1.7} & 100.0 \\
 & on & 34.5\,{\scriptsize$\pm$4.0} & 40.9\,{\scriptsize$\pm$2.0} & 79.8\,{\scriptsize$\pm$1.8} & 90.9 \\
\bottomrule
\end{tabular}

\end{table}

\section{Reasoning-mode probe}
\label{app:reasoning-probe}

We run a small reasoning-mode probe on the balanced $24$-scenario
subset of delayed-effect patches used during prompt development. It has
four scenarios per service and $960$ predictions per configuration
($24 \times 40$). It is a preliminary, non-representative check of
whether explicit reasoning changes behavior on delayed effects; it is
not part of the Core-v1 scorecard, not a substitute for Core-v1, and not
a basis for model comparison. Opus~4.7 results are reused from an earlier run on the
same subset.

\begin{table}[ht]
\centering\footnotesize
\caption{Reasoning-mode probe on $24$ delayed-effect scenarios, answers
revealed (\%). Eff.\,VM is computed on the $37$ effect steps; calls that
returned a provider error are not scored. Reasoning
tokens per call are as reported by the provider (--: not exposed). WF
is the share of parseable JSON responses. Only DeepSeek~V4-Flash exposes
verifiable reasoning-off and reasoning-on modes, so its two rows form
the controlled pair. The Sonnet~4.6 default reports no reasoning-token
count, so the two Sonnet rows are not a controlled pair; Kimi~K2.5
returned no reasoning tokens when reasoning was requested, so it has no
reasoning-on row; Opus~4.7 is a reference row. Qwen3.6-35B-A3B is
evaluated with reasoning off only.}
\label{tab:reasoning-probe}
\begin{tabular}{l|l|rrrrr}
\toprule
Configuration & Reasoning & Eff.\,VM & VM & SM & WF & Reas.\ tok/call \\
\midrule
\multicolumn{7}{l}{\emph{Controlled pair}} \\
DeepSeek V4-Flash & off & 54.1 & 81.7 & 95.1 & 100.0 & 0 \\
DeepSeek V4-Flash & on & 78.4 & 88.0 & 93.8 & 95.3 & 2567 \\
\midrule
\multicolumn{7}{l}{\emph{Unmatched rows}} \\
Claude Sonnet 4.6 & not requested & 67.6 & 75.8 & 86.2 & 89.0 & -- \\
Claude Sonnet 4.6 & requested & 73.0 & 86.5 & 91.1 & 92.9 & 104 \\
Kimi K2.5 & off & 56.8 & 81.3 & 94.9 & 100.0 & -- \\
Claude Opus 4.7 & not requested & 67.6 & 87.7 & 95.3 & 99.4 & -- \\
\bottomrule
\end{tabular}

\end{table}

\section{Free-rollout survival values}
\label{app:rollout-survival}

Table~\ref{tab:rollout-survival} reports selected points of the curve in
Figure~\ref{fig:rollout-survival}.

\begin{table}[ht]
\centering\footnotesize
\caption{Share of scenarios (\%) whose first $k$ free-rollout
responses are all exact.}
\label{tab:rollout-survival}
\begin{tabular}{l|rrrrr}
\toprule
Model & k=1 & k=5 & k=10 & k=20 & k=40 \\
\midrule
DeepSeek V4-Flash & 73.7 & 25.5 & 9.8 & 2.7 & 1.2 \\
Kimi K2.5 & 65.1 & 22.7 & 6.7 & 2.4 & 0.0 \\
Qwen3.6-35B-A3B & 66.3 & 12.2 & 3.9 & 0.4 & 0.0 \\
Claude Sonnet 4.6 & 53.7 & 12.9 & 3.1 & 0.0 & 0.0 \\
\bottomrule
\end{tabular}
\end{table}

\section{Outcome categories and parse failures}
\label{app:error-modes}

Table~\ref{tab:error-modes} splits outcomes into mutually exclusive
categories, separating prediction errors from parse failures. Across
all three protocols, at most $0.5\%$ of the responses from DeepSeek,
Kimi, and Qwen are unparseable. For Sonnet, $14.3\%$ are unparseable
with answers revealed, $13.5\%$ with no feedback, and $16.7\%$ in free
rollout; its parse rate
is thus similar across protocols and does not explain its drop between
them.

\begin{table}[ht]
\centering\footnotesize
\caption{Core-v1 outcome categories for selected model--protocol
pairs (\% of all future calls). Parse fail, VM exact, status-only, and
status wrong are mutually exclusive; WF is the share of parseable JSON
responses. ``Status-only'' is valid JSON with the correct success
or error status but a wrong payload; ``status wrong'' is valid JSON with
the wrong status.}
\label{tab:error-modes}
\begin{tabular}{l|rrrrr}
\toprule
Model, protocol & WF & Parse fail & VM exact & Status-only & Status wrong \\
\midrule
DeepSeek, answers revealed & 100.0 & 0.0 & 75.1 & 17.7 & 7.2 \\
DeepSeek, no feedback & 100.0 & 0.0 & 39.8 & 45.7 & 14.5 \\
DeepSeek, free rollout & 100.0 & 0.0 & 57.5 & 30.5 & 12.0 \\
Kimi, no feedback & 99.6 & 0.4 & 38.6 & 45.5 & 15.5 \\
Qwen, no feedback & 99.9 & 0.1 & 34.9 & 46.1 & 19.0 \\
Sonnet, no feedback & 86.5 & 13.5 & 29.5 & 45.4 & 11.5 \\
\bottomrule
\end{tabular}
\end{table}

\section{Position of the first wrong prediction}
\label{app:first-div}

For each model and protocol we record the first future call, counted
from $1$, whose prediction is wrong. Under all three protocols and for
all four models, the most common position is call $1$, because the
first future call is predicted under the same conditions in every
protocol. A first error after call $20$ is rare: it occurs in at most
$4$ of $255$ scenarios per model with answers revealed, at most $1$ with
no feedback, and at most $6$ in free rollout. The protocols differ
mainly in what happens after the first error: with answers revealed
the next prompt again contains correct responses, whereas in free
rollout the error stays in the context. Free-rollout means and medians
are in Table~\ref{tab:rollout-summary}, where an unparseable response
counts as a wrong prediction. Counting only parseable responses gives
medians of $3$/$2$/$2$/$3$ (DeepSeek/Kimi/Qwen/Sonnet); only Sonnet's
median changes, because $16.7\%$ of its free-rollout responses are
unparseable.

\section{Source-diff localization prompt}
\label{app:expl-prompt}

The source-diff localization diagnostic sends the following prompt.
Both sources are normalized so that line~1 is the first non-empty line,
and they are shown with explicit line numbers. The prompt is built in
\texttt{scripts/eval\_llm\_counterfactual.py}, function
\texttt{make\_explain\_prompt}.

\begin{lstlisting}[basicstyle=\ttfamily\footnotesize,frame=single]
Two versions of a Python service were shown to a model.
One single line of the source was changed between versions.
Name the changed line number from the line-numbered PATCHED SOURCE.

ORIGINAL SOURCE:
{numbered_original_source}

PATCHED SOURCE:
{numbered_patched_source}

Respond with ONLY a JSON object: {"line_number": N}
\end{lstlisting}

The model sees both versions of the source and is asked for the line
number of the one line that differs. This is therefore a code-diff
task, not attribution of a behavior change to a source line. A later
version should withhold the edited source and ask the model to locate
the edit from the change in responses alone.

\section{Source-diff localization: line-number audit}
\label{app:expl-audit}

The first pilot prompt inserted the service sources verbatim. Those
strings began with a blank line, so a model that counted the first
visible code line as line~1 was off by one relative to the stored
label. The released Core-v1 data strips the leading blank line, and the
evaluator displays explicit line numbers. Because the prompt contains
only the two sources, scenarios that share a patch receive the same
prompt; we therefore report exact match averaged over the $49$ unique
patches. A one-line tolerance is kept as an audit number; a two-line
tolerance is not used because several candidate edits are only a few
lines apart. These numbers are not part of the main scorecard.

\begin{table}[ht]
\centering\footnotesize
\caption{Source-diff localization diagnostic (\%). The prompt shows
both the original and the edited source, so this measures locating a
code difference, not attributing a behavior change to a line. Scores
are averaged over the $49$ unique patches; $\pm$ is the half-width of a
$95\%$ bootstrap interval over patches.}
\label{tab:source-diff}
\begin{tabular}{l|rr}
\toprule
Model & Exact line & Within one line \\
\midrule
DeepSeek V4-Flash & 7.4\,{\scriptsize$\pm$4.7} & 19.9\,{\scriptsize$\pm$8.5} \\
Kimi K2.5 & 7.3\,{\scriptsize$\pm$4.2} & 16.4\,{\scriptsize$\pm$6.6} \\
Claude Sonnet 4.6 & 54.8\,{\scriptsize$\pm$12.9} & 91.2\,{\scriptsize$\pm$6.6} \\
\bottomrule
\end{tabular}
\end{table}

\section{State-probe design note}
\label{app:sp-v2}

The version-1 state probe lets a scenario author declare the set of
\texttt{state\_dict()} paths to evaluate. A model can earn credit by copying observed values onto those paths,
because the probe cannot tell a predicted update from a copied value. We therefore neither release nor
report state-probe v1 as a Core-v1 score. A later version should change
the design in two ways.

\emph{Predeclared paths.} The probe paths for a service are declared
once in each service's schema, independently of the scenario. This removes
scenario-level degrees of freedom.

\emph{No-intervention baseline.} For every probe, compute the value
the probed field would take if the intervention had no effect, by
running the future calls on $P$ from the state after the observed calls
without applying $\xi$. A probe match counts toward the v2 score only
if the predicted value differs from this no-intervention value; this
removes the copying shortcut.

\section{Scenario selection}
\label{app:filter-v2}

For every candidate scenario we store the no-intervention trace and the
target trace for the same future calls. A candidate is kept only if at
least one response differs between the two stored traces. This removes
candidates whose intervention changes hidden state without changing any response in the stored trace. The released audit records that all $255$ Core-v1
scenarios pass. The stored traces are longer than the $40$ scored
calls; in $247$ scenarios a changed response falls within the scored
calls, and in the remaining $8$ the first change comes later, so these
$8$ contribute no effect steps. Core-v1 therefore measures whether a
model predicts changes that do happen, not whether it correctly
predicts that a change has no visible effect.

\section{No-effect probe}
\label{app:no-effect}

To examine what Core-v1 excludes, we also release
\texttt{no\_effect\_v0.jsonl}, $106$ generated scenarios whose
intervention changes none of the future responses. They are not part
of Core-v1 because they come from the delayed-effect generator and do
not match Core-v1's coverage of services and interventions. They test
an important failure mode: predicting a change after seeing an edit that
has no effect on the future calls. For reporting, this probe is superseded by
the matched companions of Appendix~\ref{app:no-effect-v1}.

\begin{table}[ht]
\centering\footnotesize
\caption{No-effect probe, answers revealed (\%). A correct prediction
equals the no-intervention response. Replaying the unchanged service would score $100\%$ VM.}
\label{tab:no-effect}
\begin{tabular}{l|rrr}
\toprule
Model & Scenarios & VM & SM \\
\midrule
DeepSeek V4-Flash & 106 & 84.5 & 96.4 \\
Kimi K2.5 & 106 & 83.1 & 95.2 \\
Claude Sonnet 4.6 & 106 & 82.3 & 93.7 \\
\bottomrule
\end{tabular}

\end{table}

\section{Matched no-effect companions (v1)}
\label{app:no-effect-v1}

Core-v1 keeps only scenarios whose intervention changes some future
response (Appendix~\ref{app:filter-v2}), so effect-step VM never
penalizes a model for predicting a change that does not happen. The
release therefore adds \texttt{counterfactuals/no\_effect\_v1.jsonl}:
$255$ no-effect companions in one-to-one correspondence with the $255$
Core-v1 scenarios (field \texttt{paired\_item\_id}).

\paragraph{Construction.} Each companion has the same service, the same
observed calls, and the same $40$ scored future calls as its Core-v1
partner, but a different intervention: a source edit or state overwrite
that changes none of the stored future responses. Admission inverts the
selection rule of Appendix~\ref{app:filter-v2}, and the check covers the
whole stored trace ($44$ to $95$ future calls), not only the $40$ scored
calls. Every companion was re-executed from its stored fields; at all
$10{,}200$ scored positions the no-intervention and target responses are
identical. Interventions are described with the same templates as in
Core-v1, and an automatic check confirms that no description asserts a
downstream effect. The split is generated deterministically by
\texttt{scripts/build\_no\_effect\_v1.py} from Core-v1 and the released
services, and rerunning it reproduces the released file byte for byte.

\paragraph{No-effect VM.} No-effect VM (NE VM) is exact value match over
all $10{,}200$ scored companion predictions. Because the target is always
the no-intervention response, every well-formed wrong prediction is an
invented change; the over-prediction rate is the share of well-formed
predictions that are wrong. Replaying the service without the
intervention scores $100\%$ NE VM by construction and $0\%$ effect-step
VM (Table~\ref{tab:baselines}).

\paragraph{Coverage.} Table~\ref{tab:ne-v1-coverage} compares the
companions and the v0 probe with Core-v1. Each companion is matched to
its Core-v1 partner on service, observed prefix, and call schedule, so
the two splits have identical distributions of service, history length,
and trajectory. The companion's intervention is chosen to be benign and
need not have the partner's type. For none of the $30$ clock-jump scenarios did the
builder find a clock value that leaves all future responses unchanged,
so these companions carry source edits ($21$ threshold changes and $9$
branch inversions). For $63$ branch-inversion scenarios it found no
benign branch inversion, and the companion carries a threshold change.
In total, $162$ companions share their partner's intervention type and
$93$ do not (field \texttt{benign\_tier}). Per-type companion scores
therefore describe the companions only and should not be compared with
Core-v1 per-type scores. Two further fields support stratified analysis.
For $114$ companions (\texttt{contingency\_class} $=$
\texttt{contingent}), the same edit changes some response on another
Core-v1 trajectory of the same service, so the companion cannot be
solved by recognizing an edit that never matters. The other $141$
(\texttt{global\_no\_op}) change no response on any of the $30$ Core-v1
trajectories of their service. In $10$ companions
(\texttt{latent\_divergence}) the edit changes hidden state but no
response.

\begin{table}[ht]
\centering\footnotesize
\caption{Coverage of the no-effect splits relative to Core-v1. Left:
total-variation distance (TVD) between scenario distributions; $0$ is an
exact match. History length groups observed histories of $5$, $15$,
$35$, and $55$ calls (state edits apply one call later and are counted in
their nominal group). A trajectory is a service with its full call
sequence; v0 scenarios cannot be mapped to Core-v1 trajectories. Right:
scenarios by intervention type.}
\label{tab:ne-v1-coverage}
\begin{tabular}{l|rr||l|rr}
\toprule
\multicolumn{3}{c||}{TVD to Core-v1} & \multicolumn{3}{c}{Scenarios by intervention type} \\
Axis & v1 & v0 & Type & Core-v1 & v1 \\
\midrule
Service & 0.000 & 0.341 & Branch inversion & 150 & 96 \\
History length & 0.000 & 0.812 & Threshold change & 60 & 144 \\
Trajectory & 0.000 & -- & Clock jump & 30 & 0 \\
Intervention type & 0.329 & 1.000 & Counter overwrite & 15 & 15 \\
\bottomrule
\end{tabular}

\end{table}

\paragraph{Results.} Table~\ref{tab:ne-v1-paired} reports paired
scores for DeepSeek~V4-Flash and Kimi~K2.5 under answers revealed. On
the Core-v1 steps whose response the intervention does not
change, the same models reach $79.6\%$ (DeepSeek) and $75.3\%$ (Kimi)
VM, in line with their NE VM.

\begin{table}[ht]
\centering\footnotesize
\caption{Paired report, answers revealed (\%). Eff.\,VM is computed on
the Core-v1 effect steps and NE VM on the $10{,}200$ companion
predictions; $\pm$ is the half-width of a $95\%$ bootstrap interval over
scenarios. Over-pred.\ is the share of well-formed companion predictions
that differ from the unchanged response. The last two columns split NE
VM into the $114$ contingent and $141$ global no-op companions.
Core-v1 runs: April 2026; companion runs: July 2026; same model
identifiers and serving routes.}
\label{tab:ne-v1-paired}
\begin{tabular}{l|cc|rrr}
\toprule
 & Eff.\,VM & NE VM & Over- & \multicolumn{2}{c}{NE VM by subset} \\
Model & (Core-v1) & (companions) & pred. & Contingent & Global no-op \\
\midrule
DeepSeek V4-Flash & 61.2\,{\scriptsize$\pm$3.1} & 78.8\,{\scriptsize$\pm$1.3} & 21.2 & 75.2 & 81.7 \\
Kimi K2.5 & 61.7\,{\scriptsize$\pm$3.1} & 75.0\,{\scriptsize$\pm$1.3} & 24.7 & 72.8 & 76.7 \\
\bottomrule
\end{tabular}

\end{table}

\paragraph{Reading the pair.} Report both coordinates. Neither model is
near a ceiling: each invents a change in more than a fifth of its
well-formed companion predictions. The pair should not be reduced to one
number. Its minimum equals effect-step VM for both models, so a minimum
would discard the no-effect axis, and the two coordinates order the
models differently: Kimi is $0.5$ points higher on effect-step VM, well
within the intervals, while DeepSeek is $3.8$ points higher on NE VM.
Both models score lower on contingent than on global no-op companions,
by $6.5$ points (DeepSeek) and $3.9$ points (Kimi). This is consistent
with models partly recognizing edits that never matter rather than
simulating the history, so the contingent subset is the stricter test.
Contingency is correlated with service (all \texttt{auth} companions are
global no-ops and all \texttt{filesystem} companions are contingent),
but in each of the four services that contain both classes the direction
is the same for both models.

\paragraph{Why NE VM must not be read alone.} NE VM can be raised by
ignoring interventions. With answers revealed, $34.3\%$ (DeepSeek) and
$28.4\%$ (Kimi) of the wrong Core-v1 effect-step predictions equal the
no-intervention response under the grader's equality test. Such a null retreat costs effect-step VM but
is rewarded on the companions. A high NE VM is therefore informative
only next to effect-step VM for the same model, protocol, and prompt,
which is why the release requires paired reporting.

\section{Identifier-renamed variant}
\label{app:renamed}

Three services port algorithms from python-limits, PyFilesystem2, and
django-oscar, so a model could in part rely on familiar library names.
The release therefore includes \texttt{counterfactuals/core\_v1\_renamed.jsonl},
a renamed copy of the $210$ source-edit scenarios (field
\texttt{paired\_item\_id}). \texttt{scripts/build\_renamed\_core.py}
rewrites, with Python's tokenizer, the service class, every operation,
helper methods, and instance attributes to neutral labels
(\texttt{SvcAUT}, \texttt{op00}, \texttt{st00}); local variables, most
parameters, comments, string literals, response keys, and error strings
are unchanged, and every edited line keeps its line number. State edits
carry no source patch and are not part of the variant. Replaying every
scenario under the renamed and the original source gives identical
responses at all $13{,}940$ observed and future calls; this is trace
verification, not a proof of equivalence. Rerunning the builder
reproduces the released file byte for byte, and
\texttt{metadata/core\_v1\_renamed\_verification.json} records the check.

We ran Kimi~K2.5 four times under answers revealed on the same day: twice
on the original $210$ scenarios and twice on the renamed copies
(Table~\ref{tab:renamed}; $202$ scenarios with $2{,}040$ effect steps, excluding calls that
returned a provider error).
Renaming lowers effect-step VM by $2.7$ points and lowers the agreement
between two runs by $4.8$ points, whereas agreement between an original
and a renamed run is as high as agreement between two renamed runs.
Renaming removes library-specific names but also generic cues such as
\texttt{withdraw}, so these differences measure reliance on identifier
meaning in general rather than recall of a particular library.
\texttt{scripts/score\_renamed\_variant.py} reproduces the table from the
released outputs.

\begin{table}[ht]
\centering\footnotesize
\caption{Identifier-renamed variant: four same-day Kimi~K2.5 runs,
answers revealed. Agreement is exact equality of two runs' predictions
on effect steps; the cross value averages the four original--renamed
pairs. Intervals are $95\%$ bootstrap intervals over scenarios;
differences resample both sides jointly.}
\label{tab:renamed}
\begin{tabular}{l|rr}
\toprule
Quantity & Value (\%) & 95\% interval \\
\midrule
\multicolumn{3}{l}{\emph{Effect-step VM}} \\
Original identifiers, run 1 & 64.5 & [61.6, 67.4] \\
Original identifiers, run 2 & 63.3 & [60.0, 66.6] \\
Renamed identifiers, run 1 & 60.7 & [57.6, 63.9] \\
Renamed identifiers, run 2 & 61.8 & [58.7, 65.0] \\
Renamed $-$ original (mean of two runs) & $-$2.7 & [$-$4.6, $-$0.8] \\
\midrule
\multicolumn{3}{l}{\emph{Prediction agreement on effect steps}} \\
Original vs.\ original & 65.4 & [62.5, 68.2] \\
Renamed vs.\ renamed & 60.6 & [58.1, 63.3] \\
Original vs.\ renamed (mean of four pairs) & 61.0 & [58.3, 63.5] \\
Cross $-$ renamed repeat & +0.4 & [$-$1.3, 2.0] \\
Renamed repeat $-$ original repeat & $-$4.8 & [$-$7.4, $-$2.2] \\
\bottomrule
\end{tabular}

\end{table}

\section{Evaluation card}
\label{app:eval-card}

\paragraph{Intended use.} \bench{} evaluates language models, without
tools, as text-only predictors of how deterministic stateful services
respond after a specified change, on fixed future calls. Core-v1 is
intended for locating failures, not for certifying models.

\paragraph{Not intended use.} Core-v1 does not evaluate agent policies,
is not a standalone free-rollout leaderboard, and does not show that a
model can attribute a behavior change to a source line. The source-diff
localization and state-probe diagnostics are not measures of response
prediction.

\paragraph{Supported claims.} The release supports claims about VM and
SM on Core-v1 under a named protocol; about how much scores fall when
correct earlier responses are hidden; about free rollout under the
released prompt, read through step-level VM and the survival curve
rather than whole-trace exact match alone; about invented changes,
through NE VM on the v1 no-effect companions reported next to
effect-step VM for the same model and protocol; about reasoning off
versus on for DeepSeek~V4-Flash; about descriptive per-service and
per-intervention failure profiles; and about the source-diff diagnostic
after averaging over unique patches. Read each score next to the
reference policy for its protocol (Appendix~\ref{app:reference-policies}).

\paragraph{Unsupported claims.} The release does not separate state
inference from simulation, does not measure attribution of behavior
changes to source lines, and supports no claims about other programming
languages or about reasoning modes beyond the DeepSeek~V4-Flash
comparison. It supports no
claim from NE VM alone, from a single number that combines NE VM and
effect-step VM, or from per-type companion scores compared with Core-v1
per-type scores.

\paragraph{Protocol names in the release.} The released files and
scripts use code names for the protocols: \texttt{tf1step\_sliding\_k20}
is answers revealed, \texttt{no\_suffix\_feedback} is no feedback, and
\texttt{free\_rollout} is free rollout. All scores in this paper use
prompt v1 (\texttt{-{}-prompt-version v1}); the release also provides
prompt v1.1 for new evaluations and documents both versions.

\paragraph{Reporting a new model.} State the dataset version (Core-v1),
the protocol (answers revealed, no feedback, or free rollout), the prompt
version, the
model version and access date, the serving route and reasoning setting,
the temperature, the token budget, the retry and parse policy, the
parser and grader version, the evaluated subset if it is not the full
split, and the inference cost when available. Report source-diff numbers as averages over
the $49$ unique patches. Start from the predictors in
Table~\ref{tab:baselines}. The minimum report is the pair (effect-step
VM on Core-v1, NE VM on the v1 companions) under answers revealed, each
with its bootstrap interval, as computed by
\texttt{scripts/score\_paired.py}; the pair should not be reduced to one
number (Appendix~\ref{app:no-effect-v1}). Free-rollout effect-step VM
should be added whenever a claim concerns unaided simulation.

\paragraph{Known limitations of the release.} Core-v1 contains only
Python services. Answers revealed supplies correct earlier responses,
and the window is fixed at $20$ calls without a window-size ablation.
State-probe v1 can reward copying the observed state. Source-diff
localization v1 shows both source versions and repeats prompts for
scenarios that share a patch. Core-v1 excludes scenarios in which the
intervention has no visible effect; the v1 companions cover this case
and are matched to Core-v1 on service, observed prefix, and call
schedule (Appendix~\ref{app:no-effect-v1}). Larger repositories and new
service families are outside the current evidence; human review is
suited to auditing scenario validity and service relevance, while
scoring stays automatic against the executed oracle.

\paragraph{Licenses and model outputs.} The benchmark code is released
under the MIT license and the dataset under CC~BY~4.0. The
\texttt{ratelimiter}, \texttt{filesystem}, and \texttt{reservation}
services port algorithms from python-limits (MIT), PyFilesystem2 (MIT),
and django-oscar (BSD 3-Clause); the release reproduces their license
texts and copyright notices. The released reference outputs are
AI-generated; the dataset card summarizes the providers' terms for
their reuse.

\end{document}